\documentclass[10pt]{article}
\usepackage{graphicx}
\usepackage{amsmath}
\usepackage{amssymb}
\usepackage{caption2}
\begin{document}
\begin{center}
{\large {\bf \sc{  Comment on ``Comment on ``Landau equation and QCD sum rules for the tetraquark molecular states""
  }}} \\[2mm]
Zhi-Gang  Wang \footnote{E-mail: zgwang@aliyun.com.  }  \\
 Department of Physics, North China Electric Power University, Baoding 071003, P. R. China
\end{center}

\begin{abstract}
In ``Landau equation and QCD sum rules for the tetraquark molecular states", I list out six reasons to refute the assertion of
Lucha, Melikhov and Sazdjian. In ``Comment on ``Landau equation and QCD sum rules for the tetraquark molecular states"", they refute my viewpoint. If they want to the refute my viewpoint robustly, they should refute the six reasons one by one, and provide at least one  example to illustrate that they have obtained reliable QCD sum rules for the tetraquark states and free two-meson states, respectively.
\end{abstract}

 \bigskip

In Ref.\cite{WZG-Landau}, I list out six reasons to refute the assertion of  Lucha, Melikhov and Sazdjian, they assert that the contributions  at the order $\mathcal{O}(\alpha_s^k)$ with $k\leq1$ in the operator product expansion, which are factorizable in the color space, are exactly  canceled out   by the meson-meson scattering states at the hadron side, the tetraquark (molecular) states begin to receive contributions at the order $\mathcal{O}(\alpha_s^2)$ \cite{Chu-Sheng-PRD-1,Chu-Sheng-PRD-2}. \\

Let us write down the two-point correlation functions $\Pi(p)$ in the QCD sum rules,
\begin{eqnarray}
\Pi(p)&=&i\int d^4x e^{ip\cdot x}\langle 0|T\left\{ J(x)J^\dagger(0)\right\}|0\rangle\, ,
\end{eqnarray}
where the $J(x)$ are the tetraquark currents. In the operator product expansion, the Feynman diagrams can be divided into both factorizable and nonfactorizable diagrams in the color space, while they are all nonfactorizable in the momentum space.
At the hadron side of the QCD sum rules, I use the Ansatz  ``{\bf meson-meson scattering states} plus tetraquark (molecular) states", according to the nonfactorizable properties in the momentum space.
  While  Lucha, Melikhov and Sazdjian use the Ansatz  ``{\bf free two-meson states} plus {\bf two-meson scattering states} plus tetraquark (molecular) states" \cite{Chu-Sheng-PRD-1}.
The term ``{\bf free two-meson states}" is not robust, as there are momentum flows  in all the Feynman diagrams. 
Roughly speaking, my term ``{\bf meson-meson scattering states}" equals their term  ``{\bf free two-meson states} plus {\bf two-meson scattering states}";
at the  contributions  of the order $\mathcal{O}(\alpha_s^k)$ with $k\leq1$, my term ``{\bf meson-meson scattering states}" equals their term  ``{\bf free two-meson states}". All in all, I do not misunderstand the essence of their assertion that the tetraquark (molecular) states begin to receive contributions at the order $\mathcal{O}(\alpha_s^2)$.   \\

In the following, I list out the six reasons raised in Ref.\cite{WZG-Landau} one by one, for details, one can consult Ref.\cite{WZG-Landau}. \\

{\bf 1.}  We cannot assert that the factorizable  Feynman diagrams in color space are exactly canceled out by the meson-meson scattering states, because the meson-meson scattering state and tetraquark molecular state both have four valence quarks, which can be divided into  two color-neutral clusters. We cannot distinguish which Feynman diagrams contribute to the  meson-meson scattering state or tetraquark molecular state based on the two color-neutral clusters.

In the comment \cite{ChuSheng-2005}, Lucha, Melikhov and Sazdjian emphasize that the contributions  at the order $\mathcal{O}(\alpha_s^k)$ with $k\leq1$ in the operator product expansion are exactly  canceled out   by the {\bf free two-meson states} at the hadron side. The arguments they presented in  Ref.\cite{ChuSheng-2005} are just  repetitions of the old arguments presented in Refs.\cite{Chu-Sheng-PRD-1,Chu-Sheng-PRD-2}. If they want to prove that their viewpoint is right,  they should provide an example to show that   the hadron side of the QCD sum rules  can be saturated by the  {\bf free two-meson states} indeed, just like the QCD sum rules in Eqs.(32-35) in Ref.\cite{WZG-Landau}. Only arguments, it is vague.
In fact, it is of no use  to distinguish the factorizable and nonfactorizable properties of the Feynman diagrams,  we can only obtain information about the short-distance and long-distance contributions in the operator product expansion. \\

{\bf 2.} The quarks and gluons are confined objects, they cannot be put on the mass-shell.
If we insist on applying  the Landau equation to study the Feynman diagrams, we should choose the pole masses rather than the $\overline{MS}$ masses, which lead to obvious problems in the QCD sum rules for the traditional  charmonium and bottomonium states.

In the comment \cite{ChuSheng-2005}, Lucha, Melikhov and Sazdjian do not touch this subject. \\

{\bf 3.} The nonfactorizable Feynman diagrams in the color space which have the Landau singularities begin to appear at the order $\mathcal{O}(\alpha_s^0/\alpha_s^1)$ rather than at the order $\mathcal{O}(\alpha_s^2)$ due to the nonperturvative contributions, again if we insist on applying  the Landau equation to study the Feynman diagrams.

In the comment \cite{ChuSheng-2005}, Lucha, Melikhov and Sazdjian do not touch the contributions of the Feynman diagrams involving the vacuum condensates, see Fig.2 in Ref.\cite{WZG-Landau}. \\

{\bf 4.}  Without taking it for granted that the  factorizable Feynman diagrams in the  color space only make contributions to the meson-meson scattering states,  the Landau equation cannot exclude the factorizable Feynman diagrams in the color space, those diagrams can also have the Landau singularities and make contributions to the tetraquark (molecular) states.

In the comment \cite{ChuSheng-2005}, Lucha, Melikhov and Sazdjian just argue that if such diagrams make contributions to the tetraquark (molecular) states, they are in contradiction with the Large-$N_c$ behavior of the QCD Green functions. In reality, $N_c=3$, we should be care to apply the  Large-$N_c$ arguments in the QCD sum rules naively.
In Ref.\cite{Chu-Sheng-PRD-2}, Lucha, Melikhov and Sazdjian use a triplet-antitriplet tetraquark current (also an octet-octet current),
\begin{eqnarray}
(\epsilon^{ijk}\bar{q}_a^j\bar{q}^k_c )(\epsilon^{ij^\prime k^\prime}q_b^{j^\prime}q^{k^\prime}_d )&=&-\theta_{\bar{a}b\bar{c}d}-\theta_{\bar{a}d\bar{c}b}\, ,
\end{eqnarray}
to illustrate that it is  sufficient to study the QCD sum rules for the exotic interpolating currents taken as products
of two color-singlet quark bilinears $\theta_{\bar{a}b\bar{c}d}$ and $\theta_{\bar{a}d\bar{c}b}$.
If we take the   Large-$N_c$ limit, the tetraquark current should be modified to be  a 2$(N_c-1)$-quark current,
\begin{eqnarray}
(\epsilon^{ii_1i_2\cdots i_{N_c-1}}\bar{q}_{a_1}^{i_1}\bar{q}^{i_2}_{a_2}\cdots \bar{q}_{a_{N_c-1}}^{i_{N_c-1}} )(\epsilon^{ij_1j_2\cdots j_{N_c-1}}q_{b_1}^{j_1}q^{j_2}_{b_2}\cdots q_{b_{N_c-1}}^{j_{N_c-1}} )&=&\bar{q}_{a_1}q_{b_1}\bar{q}_{a_2}q_{b_2}\cdots \bar{q}_{a_{N_c-1}}q_{b_{N_c-1}}\nonumber\\
&&+\cdots \, ,
\end{eqnarray}
where the $a_1$, $b_1$, $\cdots$, $a_{N_c-1}$ and $b_{N_c-1}$ are the flavor indexes.
We should choose the currents $\bar{q}_{a_1}q_{b_1}\bar{q}_{a_2}q_{b_2}\cdots \bar{q}_{a_{N_c-1}}q_{b_{N_c-1}}$, $\cdots$ rather than the bilinears $\theta_{\bar{a}b\bar{c}d}$ and $\theta_{\bar{a}d\bar{c}b}$ to discuss  the large-$N_c$ behavior. The currents $\bar{q}_{a_1}q_{b_1}\bar{q}_{a_2}q_{b_2}\cdots \bar{q}_{a_{N_c-1}}q_{b_{N_c-1}}$, $\cdots$ couple to the 2$(N_c-1)$-quark states or free $(N_c-1)$-meson states with masses of the order $\mathcal{O}(N_c)$.
Furthermore, in the QCD sum rules, we do not resort to the parameter $\frac{1}{N_c}$ in the operator product expansion.     \\

{\bf 5.}  Lucha,  Melikhov and  Sazdjian only obtain formal QCD sum rules for the tetraquark  (molecular) states.

In the comment \cite{ChuSheng-2005}, Lucha, Melikhov and Sazdjian do not touch this subject. If they want to illustrate that they can obtain reliable QCD sum rules for the tetraquark states without contaminations, they should provide an example, just like the QCD sum rules in Eqs.(36-39) in Ref.\cite{WZG-Landau}. If they can provide an example with predictions for the tetraquark masses, the basic parameters for the tetraquark states,  I can admit their arguments are right.   In Ref.\cite{WZG-Landau}, we choose two currents to illustrate that we can obtain the reliable QCD sum rules for the tetraquark (molecular) masses.   \\

{\bf 6.}  In the QCD sum rules, we carry out the operator product expansion in  the deep Euclidean space, $-p^2 \to \infty \,\,\,{\rm or}\,\gg \Lambda_{QCD}^2$. The Landau singularities require that the squared momentum $p^2=(\hat{m}_a+\hat{m}_b+\hat{m}_c+\hat{m}_d)^2$ in the Feynman diagrams, it is questionable  to perform the operator product expansion.

In the comment \cite{ChuSheng-2005}, Lucha, Melikhov and Sazdjian do not touch this subject. \\

In the conclusion of the comment \cite{ChuSheng-2005},  Lucha, Melikhov and Sazdjian say ``the Ansatz for the phenomenological side of
a QCD sum rule in the form ``the tetraquark pole plus the effective continuum equal to the pQCD diagrams above some effective threshold"". It is wrong, in the QCD sum rules for the tetraquark states, we use the $\rho_{QCD}(s)\Theta(s-s_0)$ to approximate the continuum states, the $\rho_{QCD}(s)\Theta(s-s_0)$ receive both short-distance and long-distance contributions, where the $s_0$ are the continuum threshold parameters.

Also, in the conclusion of the comment \cite{ChuSheng-2005},  Lucha, Melikhov and Sazdjian say ``the criterion of selecting the tetraquark-relevant
diagrams on the basis of their four-quark singularities applies to the four-point functions, and not to the
two-point function". I want to say that in the QCD sum rules, we resort to the two-point functions rather than the four-point functions to study the hadron masses. They can provide at least one  example to illustrate that they have obtained reliable QCD sum rules for the tetraquark states and free two-meson states, respectively, based on the two-point functions.

\section*{Acknowledgements}
This  work is supported by National Natural Science Foundation, Grant Number  11775079.

\end{document}